\documentclass[12pt]{article}
\usepackage{graphicx}

\begin{document}

\begin{center}

{\large Decoherence and Rabi oscillations in a qubit coupled to a
quantum two-level system}

\

S. Ashhab$^1$, J. R. Johansson$^1$, and Franco Nori$^{1,2}$

{\footnotesize 1) Frontier Research System, The Institute of
Physical and Chemical Research (RIKEN), Wako-shi, Saitama, Japan
\\ 2) Center for Theoretical Physics, CSCS, Department of Physics,
University of Michigan, Ann Arbor, Michigan, USA}

\vspace{1.0cm} {\bf Abstract}
\end{center}

In this paper we review some of our recent results on the problem
of a qubit coupled to a quantum two-level system. We consider both
the decoherence dynamics and the qubit's response to an
oscillating external field.

\vspace{1.5cm}

\section{Introduction}

Significant advances in the field of superconductor-based quantum
information processing have been made in recent years \cite{You1}.
However, one of the major problems that need to be treated before
a quantum computer can be realized is the problem of decoherence.
Recent experiments on the sources of qubit decoherence saw
evidence that the qubit was strongly coupled to quantum two-level
systems (TLSs) with long decoherence times \cite{Simmonds,Cooper}.
Furthermore, it is well known that the qubit decoherence dynamics
can depend on the exact nature of the noise causing that
decoherence. For example, an environment comprised of a large
number of TLSs that are all weakly coupled to the qubit will
generally cause non-Markovian decoherence dynamics in the qubit.
The two above observations comprise our main motivation to study
the decoherence dynamics of a qubit coupled to a quantum TLS.

A related problem in the context of the present study is that of
Rabi oscillations in a qubit coupled to a TLS
\cite{Cooper,Galperin,Ashhab2}. That problem is of great
importance because of the ubiquitous use of Rabi oscillations as a
qubit manipulation technique. We perform a systematic analysis
with the aim of understanding various aspects of this phenomenon
and seeking useful applications of it. Note that the results of
this analysis are also relevant to the problem of Rabi
oscillations in a qubit that is interacting with other surrounding
qubits.

This paper is organized as follows: in Sec. 2 we introduce the
model system and Hamiltonian. In Sec. 3 we analyze the problem of
qubit decoherence in the absence of an external driving field. In
Sec. 4 we discuss the Rabi-oscillation dynamics of the qubit-TLS
system. We finally conclude our discussion in Sec. 5.

\section{Model system and Hamiltonian}

The model system that we shall study in this paper is comprised of
a qubit that can generally be driven by an harmonically
oscillating external field, a quantum TLS and their weakly-coupled
environment \cite{Assumptions}. We assume that the qubit and the
TLS interact with their own (uncorrelated) environments that would
cause decoherence even in the absence of qubit-TLS coupling. The
Hamiltonian of the system is given by:

\begin{equation}
\hat{H}(t) = \hat{H}_{\rm q}(t) + \hat{H}_{\rm TLS} + \hat{H}_{\rm
I} + \hat{H}_{\rm Env},
\end{equation}

\noindent where $\hat{H}_{\rm q}$ and $\hat{H}_{\rm TLS}$ are the
qubit and TLS Hamiltonians, respectively, $\hat{H}_{\rm I}$
describes the coupling between the qubit and the TLS, and
$\hat{H}_{\rm Env}$ describes all the degrees of freedom in the
environment and their coupling to the qubit and TLS. The
(generally time-dependent) qubit Hamiltonian is given by:

\begin{equation}
\hat{H}_{\rm q}(t) = -\frac{E_{\rm q}}{2}\left(\sin \theta_{\rm
q}\hat\sigma_x^{\rm (q)}+\cos \theta_{\rm q}\hat\sigma_z^{\rm
(q)}\right)+F\cos(\omega t)\left(\sin \theta_{\rm
f}\hat\sigma_x^{\rm (q)}+\cos \theta_{\rm f}\hat\sigma_z^{\rm
(q)}\right), \label{eq:QubitHamiltonian}
\end{equation}

\noindent where $E_{\rm q}$ and $\theta_{\rm q}$ are the
adjustable control parameters of the qubit,
$\hat{\sigma}_{\alpha}^{\rm (q)}$ are the Pauli spin matrices of
the qubit, $F$ and $\omega$ are the amplitude (in energy units)
and frequency, respectively, of the driving field, and
$\theta_{\rm f}$ is an angle that describes the orientation of the
external field relative to the qubit $\hat\sigma_z$ axis. We
assume that the TLS is not coupled to the external driving field,
and its Hamiltonian is given by:

\begin{equation}
\hat{H}_{\rm TLS} = -\frac{E_{\rm TLS}}{2}\left(\sin \theta_{\rm
TLS}\hat\sigma_x^{\rm (TLS)}+\cos \theta_{\rm
TLS}\hat\sigma_z^{\rm (TLS)}\right),
\end{equation}

\noindent where the parameters and operators are defined similarly
to those of the qubit, except that the parameters are
uncontrollable. Note that our assumption that the TLS is not
coupled to the driving field can be valid even in cases where the
physical nature of the TLS and the driving field leads to such
coupling, since we generally consider a microscopic TLS, rendering
any coupling to the external field negligible. The qubit-TLS
interaction Hamiltonian is given by:

\begin{equation}
\hat{H}_{\rm I} = - \frac{\lambda}{2} \hat{\sigma}_z^{\rm (q)}
\otimes \hat{\sigma}_z^{\rm (TLS)},
\end{equation}

\noindent where $\lambda$ is the (uncontrollable) qubit-TLS
coupling strength. Note that, with an appropriate basis
transformation, this is a rather general form for $\hat H_{\rm I}$
\cite{Assumptions}.

\section{Qubit decoherence in the absence of a driving field}

We start by studying the effects of a single quantum TLS on the
qubit decoherence dynamics. We shall assume that all the coupling
terms in $\hat{H}_{\rm Env}$ are small enough that its effect on
the dynamics of the qubit+TLS system can be treated within the
framework of the Markovian Bloch-Redfield master equation
approach. The quantity that we need to study is therefore the 4
$\times$ 4 density matrix describing the qubit-TLS combined
system. Following the standard procedure, as can be found in Ref.
\cite{Cohen_Tannoudji}, we can write a master equation that
describes the time-evolution of that density matrix. We shall not
include that master equation explicitly here (see Ref.
\cite{Ashhab1}). Once we find the dynamics of the combined system,
we can trace out the TLS degree of freedom to find the dynamics of
the reduced $2 \times 2$ density matrix describing the qubit
alone. From that dynamics we can infer the effect of the TLS on
the qubit decoherence and, whenever the decay can be fit well by
exponential functions, extract the qubit dephasing and relaxation
rates.

Since we shall consider in some detail the case of a weakly
coupled TLS, and we shall use numerical calculations as part of
our analysis, one may ask why we do not simulate the decoherence
dynamics of a qubit coupled to a large number of such TLSs.
Focussing on one TLS has the advantage that we can obtain analytic
results describing the contribution of that TLS to the qubit
decoherence. That analysis can be more helpful in building an
intuitive understanding of the effects of an environment comprised
of a large number of TLSs than a more sophisticated simulation of
an environment comprised of, say, twenty TLSs. The main purpose of
using the numerical simulations in this work is to study the
conditions of validity of our analytically obtained results.

\subsection{Analytic results for the weak-coupling limit}

If we take the limit where $\lambda$ is much smaller than any
other energy scale in the problem \cite{Temperature}, and we take
the TLS decoherence rates to be substantially larger than those of
the qubit, we can perform a perturbative calculation on the master
equation and obtain analytic expressions for the TLS contribution
to the qubit decoherence dynamics. If we take the above-mentioned
limit and look for exponentially decaying solutions with rates
that approach the unperturbed relaxation and dephasing rates
$\Gamma_1^{\rm (q)}$ and $\Gamma_2^{\rm (q)}$, we find the
following approximate expressions for the leading-order
corrections (we take $\hbar=1$):

\noindent
\parbox{11cm}{
\begin{eqnarray*}
\delta \Gamma_1^{\rm (q)} & \approx & \frac{1}{2} \lambda^2 \sin^2
\theta_{\rm q} \sin^2 \theta_{\rm TLS} \frac{\Gamma_2^{\rm (TLS)}
+ \Gamma_2^{\rm (q)} - \Gamma_1^{\rm (q)} }{ \left(\Gamma_2^{\rm
(TLS)}+\Gamma_2^{\rm (q)}-\Gamma_1^{\rm (q)}\right)^2 + \left(
E_{\rm q} - E_{\rm TLS} \right)^2}
\\
\delta \Gamma_2^{\rm (q)} & \approx & \frac{1}{4} \lambda^2 \sin^2
\theta_{\rm q} \sin^2 \theta_{\rm TLS} \frac{\Gamma_2^{\rm
(TLS)}-\Gamma_2^{\rm (q)} }{ \left(\Gamma_2^{\rm
(TLS)}-\Gamma_2^{\rm (q)}\right)^2 + \left( E_{\rm q} - E_{\rm
TLS} \right)^2}.
\end{eqnarray*}
} \hfill
\parbox{1cm}{\begin{equation} \label{eq:Pert_theory_rates} \end{equation}}

\noindent The above expressions can be considered a generalization
of the well-known results of the traditional weak-coupling
approximation (see e.g. Ref. \cite{Shnirman}). The two approaches
agree in the limit where they are both expected to apply very
well, namely when the decoherence times of the TLS are much
shorter than those of the qubit. We shall discuss shortly,
however, that our expressions have a wider range of validity.

\subsection{Numerical solution of the master equation}

Given the large number of parameters that can be varied, we
restrict ourselves to certain special cases that we find most
interesting to analyze \cite{Temperature}. Since the TLS effects
on the qubit dynamics are largest when the two are resonant with
each other, we set $E_{\rm q}=E_{\rm TLS}$. Furthermore, we are
assuming that the energy splitting, which is the largest energy
scale in the problem, to be much larger than all other energy
scales, such that its exact value does not affect any of our
results. We are therefore left with the background decoherence
rates and the coupling strength as free parameters that we can
vary in order to study the different possible types of behaviour
in the qubit dynamics.

We first consider the weak-coupling regimes. Characterizing the
dynamics is most easily done by considering the relaxation
dynamics. Figure 1 shows the relative correction to the qubit
relaxation rate as a function of time for three different sets of
parameters differing by the relation between the qubit and TLS
decoherence rates, maintaining the relation $\Gamma_2^{\rm
(q)}/\Gamma_1^{\rm (q)}=\Gamma_2^{\rm (TLS)}/\Gamma_1^{\rm
(TLS)}=2$. We can see that there are several possible types of
behaviour of the qubit dynamics depending on the choice of the
different parameters in the problem. As a general simple rule,
which is inspired by Fig. 1(a), we find that the relaxation rate
starts at its unperturbed value and follows an exponential decay
function with a characteristic time given by $(\Gamma_2^{\rm
(TLS)}+\Gamma_2^{\rm (q)}-\Gamma_1^{\rm (q)})^{-1}$, after which
it saturates at a steady-state value given by Eq.
(\ref{eq:Pert_theory_rates}) (with $E_{\rm q}=E_{\rm TLS}$):

\noindent
\begin{equation}
\frac{dP_{\rm{ex}}(t)/dt}{P_{\rm ex}(t)-P_{\rm ex}(\infty)}
\approx - \Gamma_1^{\rm (q)} - \delta\Gamma_1^{\rm (q)} \left( 1-
\exp{\left\{- \left( \Gamma_2^{\rm (TLS)}+\Gamma_2^{\rm
(q)}-\Gamma_1^{\rm (q)} \right) t\right\}} \right).
\label{eq:General_Relaxation Rate}
\end{equation}
\noindent We can therefore say that the qubit relaxation starts
with an exponential-times-Gaussian decay function. Whether that
function holds for all relevant times or it turns into an
exponential-decay function depends on the relation between
$\Gamma_1^{\rm (q)}$ and $\Gamma_2^{\rm (TLS)}+\Gamma_2^{\rm
(q)}$. In the limit when the TLS decoherence rates are much larger
than those of the qubit, the qubit decoherence rate saturates
quickly to a value that includes the correction given in Eq.
(\ref{eq:Pert_theory_rates}). In the opposite limit, i.e. when the
TLS decoherence rates are much smaller than those of the qubit,
the contribution of the TLS to the qubit relaxation dynamics is a
Gaussian decay function. It is worth mentioning here that all the
curves shown in Fig. 1 agree very well with Eq.
(\ref{eq:General_Relaxation Rate}).

\begin{figure*}[ht]
\begin{minipage}[b]{5.3cm}
\includegraphics[width=5.3cm]{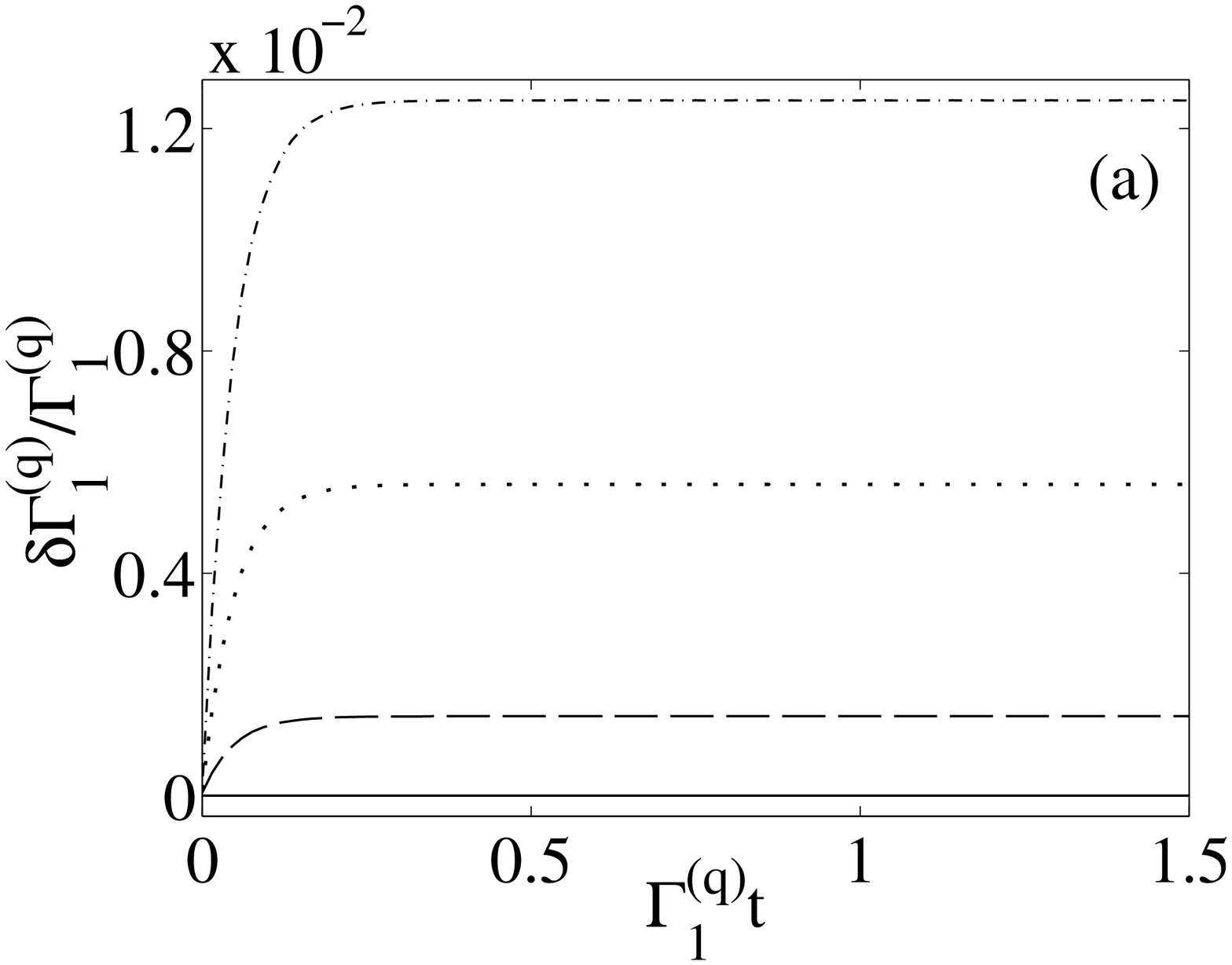}
\end{minipage}
\begin{minipage}[b]{5.2cm}
\includegraphics[width=5.2cm]{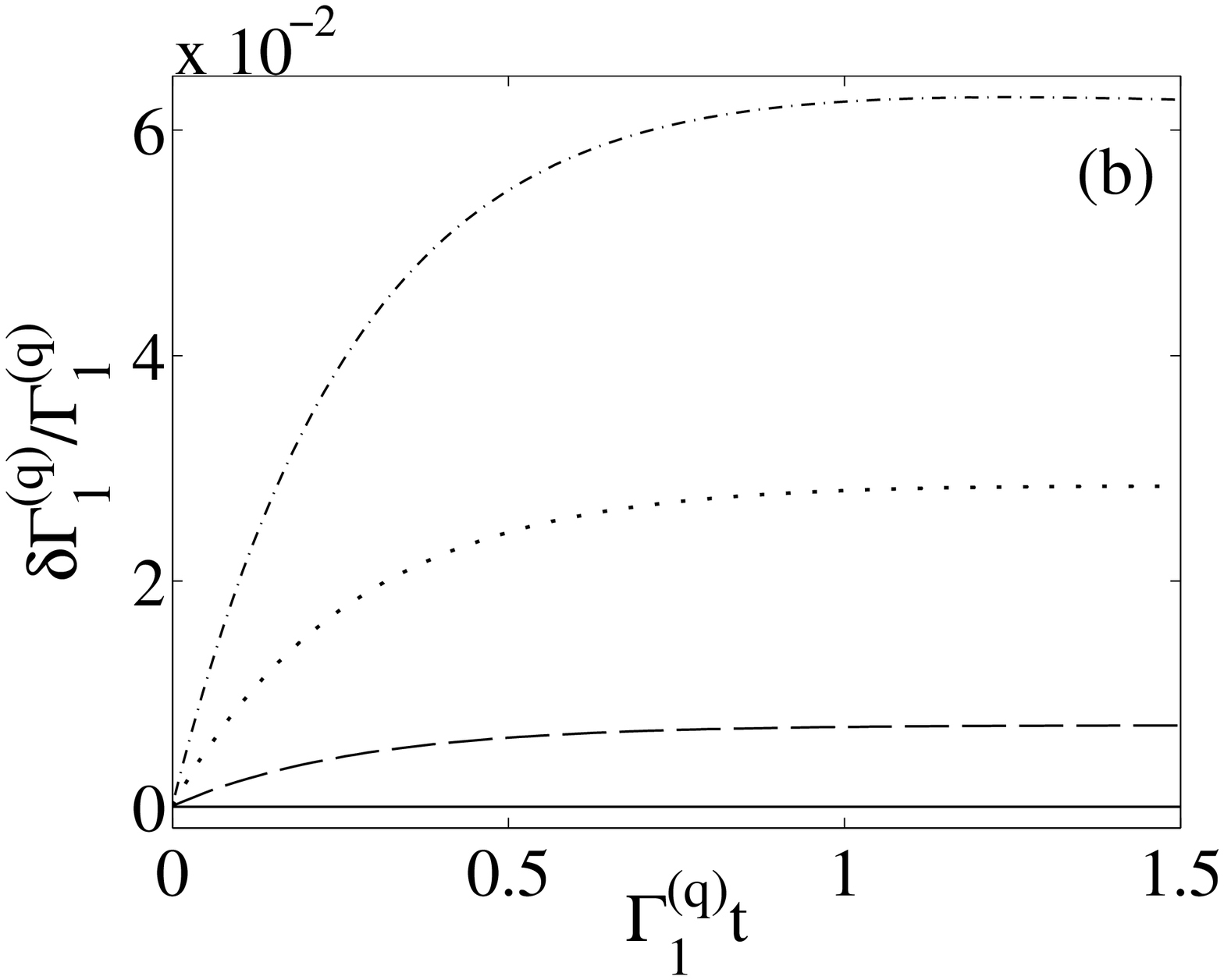}
\end{minipage}
\begin{minipage}[b]{5.3cm}
\includegraphics[width=5.3cm]{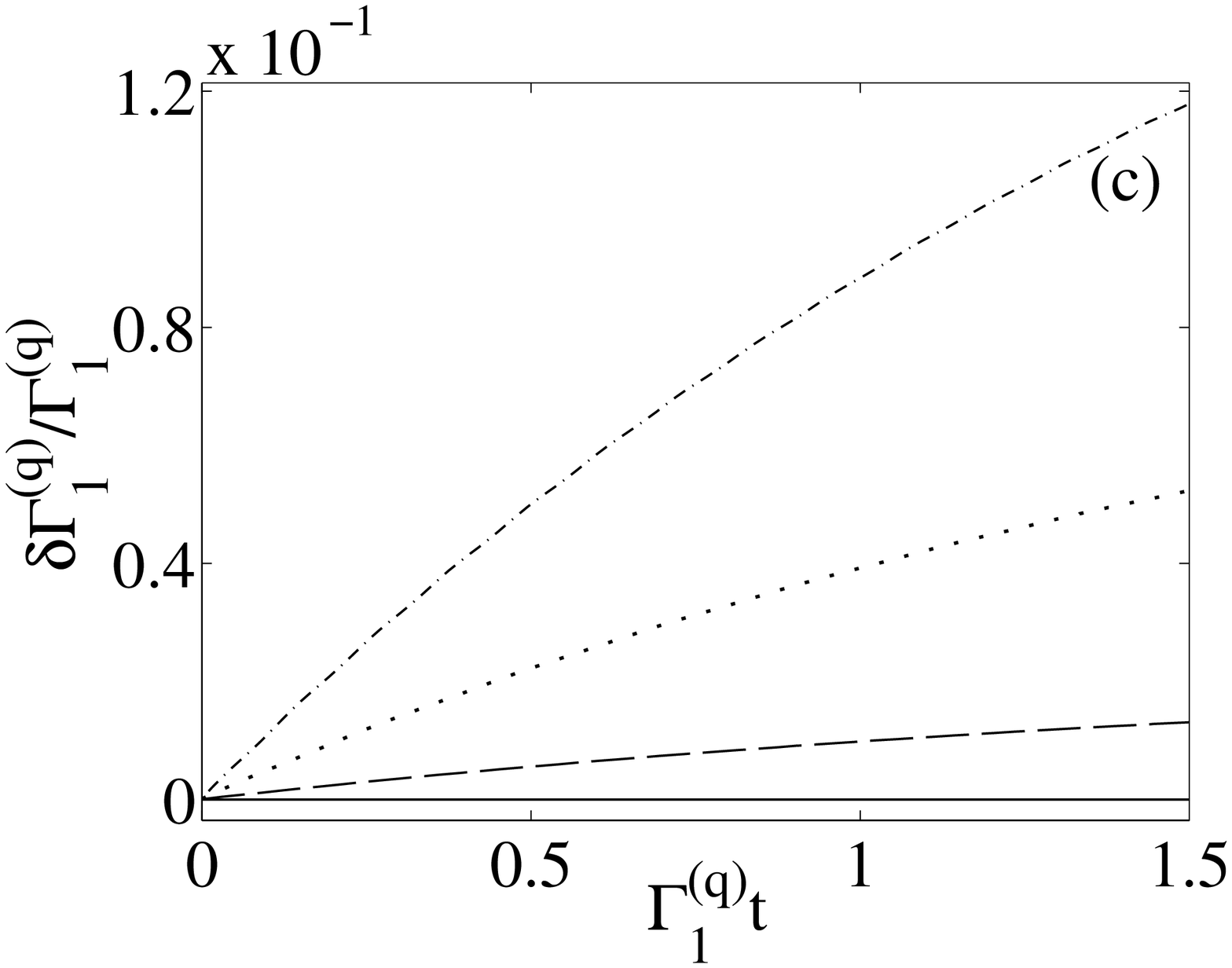}
\end{minipage}
\caption{Relative corrections to qubit relaxation rate as a
function of scaled time in the case of (a) strongly, (b)
moderately and (c) weakly dissipative TLS. The ratio
$\Gamma_1^{\rm (TLS)}/\Gamma_1^{\rm (q)}$ is 10 in (a), 1.5 in (b)
and 0.1 in (c). The solid, dashed, dotted and dash-dotted lines
correspond to $\lambda/\Gamma_1^{\rm (q)}=0$, 0.3, 0.6 and 0.9,
respectively. $\theta_{\rm q} = \pi/3$ and $\theta_{\rm
TLS}=3\pi/8$.}
\end{figure*}

The dephasing dynamics was somewhat more difficult to analyze. The
dephasing rate generally showed oscillations with frequency
$E_{\rm q}$, and the amplitude of the oscillations grew with time,
making it difficult to extract the dynamics directly from the raw
data for the dephasing rate. However, the averaged dephasing rate,
taken over one or two oscillation periods, is fit well to the
formula:

\noindent
\begin{equation}
\frac{1}{\rho_{01}}\left(\frac{\rm{d}\rho_{01}}{\rm{d}t}\right)
\approx - \Gamma_2^{\rm (q)} - \delta\Gamma_2^{\rm (q)}\left( 1-
\exp{\left\{- \left( \Gamma_2^{\rm (TLS)}-\Gamma_2^{\rm (q)}
\right) t\right\}} \right). \label{eq:General_Dephasing Rate}
\end{equation}

\noindent where $\delta\Gamma_2^{\rm (q)}$ is given by Eq.
(\ref{eq:Pert_theory_rates}) with $E_{\rm q}=E_{\rm TLS}$.

In the strong-coupling regime corresponding to large values of
$\lambda$, one cannot simply speak of a TLS contribution to qubit
decoherence. We therefore do not discuss that case here. Instead,
we discuss the transition from weak to strong coupling. We use the
criterion of visible deviations in the qubit dynamics from
exponential decay as a measure of how strongly coupled a TLS is.
The results of our calculations can be summarized as follows: a
given TLS can be considered to interact weakly with the qubit if
the coupling strength $\lambda$ is smaller than the largest
background decoherence rate in the problem. The exact location of
the boundary, however, varies by up to an order of magnitude
depending on which part of the dynamics we consider (e.g.
relaxation vs. dephasing) and how large a deviation from
exponential decay we require.

We have also checked the boundary beyond which the numerical
results disagree with our analytic expressions given in Eq.
(\ref{eq:Pert_theory_rates}), and we found that the boundary is
similar to the one given above. That result confirms the statement
made in Sec. 3.2 that our analytic expressions describing the
contribution of the TLS to the decoherence rates have a wider
range of validity than those of the traditional weak-coupling
approximation.

\section{Dynamics under the influence of a driving field}

We now include the oscillating external field in the qubit
Hamiltonian (Eq. \ref{eq:QubitHamiltonian}). Furthermore, since
decoherence does not have any qualitative effect on the main ideas
discussed here, we neglect decoherence completely in most of this
section.

\subsection{Intuitive picture}

In order for a given experimental sample to function as a qubit,
the qubit-TLS coupling strength $\lambda$ must be much smaller
than the energy splitting of the qubit $E_{\rm q}$. We therefore
take that limit and straightforwardly find the energy levels to be
given by:

\begin{eqnarray}
E_{1,4} = \mp \frac{E_{\rm TLS}+E_{\rm q}}{2} - \frac{\lambda_{\rm
cc}}{2}\, ;\,\,\,  E_{2,3} = \mp \frac{1}{2} \sqrt{ \left(E_{\rm
TLS}-E_{\rm q}\right)^2 + \lambda_{\rm ss}^2} + \frac{\lambda_{\rm
cc}}{2}, \label{eq:Energy_levels}
\end{eqnarray}

\noindent where $\lambda_{\rm cc}=\lambda \cos\theta_{\rm q}
\cos\theta_{\rm TLS}$, and $\lambda_{\rm ss}=\lambda
\sin\theta_{\rm q} \sin\theta_{\rm TLS}$.

If a qubit with energy splitting $E_{\rm q}$ is driven by a
harmonically oscillating field with a frequency $\omega$ close to
its energy splitting as described by Eq.
(\ref{eq:QubitHamiltonian}), one obtains the well-known Rabi
oscillation peak in the frequency domain with on-resonance Rabi
frequency $\Omega_0=F |\sin(\theta_{\rm f}-\theta_{\rm q})|/2$ and
full $g \leftrightarrow e$ conversion probability on resonance.
Note that the width of the Rabi peak in the frequency domain is
also given by $\Omega_0$.

Simple Rabi oscillations can also be observed in a multi-level
system if the driving frequency is on resonance with one of the
relevant energy splittings but off resonance with all others. We
can therefore combine the above arguments as follows: The driving
field tries to flip the state of the qubit alone, with a typical
time scale of $\Omega_0^{-1}$, whereas the TLS can respond to the
qubit dynamics on a time scale of $\lambda_{\rm ss}^{-1}$.
Therefore if $\Omega_0\gg\lambda_{\rm ss}$, we expect the TLS to
have a negligible effect on the Rabi oscillations. If, on the
other hand, $\Omega_0$ is comparable to or smaller than
$\lambda_{\rm ss}$, the driving field becomes a probe of the
four-level spectrum of the combined qubit-TLS system.

\subsection{Numerical results}

In order to study the Rabi-oscillation dynamics in this system, we
analyze the quantity $P_{\rm \uparrow ,max}^{(\rm q)}$, which is
defined as the maximum probability for the qubit to be found in
the excited state between times $t=0$ and $t=20\pi/\Omega_0$.
Figure 2 shows $P_{\rm \uparrow ,max}^{(\rm q)}$ as a function of
detuning ($\delta\omega\equiv\omega-E_{\rm q}$) for different
values of coupling strength $\lambda$. In addition to the
splitting of the Rabi peak into two peaks, we see an additional
sharp peak at zero detuning and some additional dips. The peak can
be explained as a two-photon transition where both qubit and TLS
are excited from their ground states to their excited states (note
that $E_{\rm q}=E_{\rm TLS}$). The dips can be explained as
``accidental'' suppressions of the oscillation amplitude when one
energy splitting in the four-level spectrum is a multiple of
another energy splitting in the spectrum.

\begin{figure}[ht]
\begin{minipage}[b]{5.3cm}
\includegraphics[width=5.3cm]{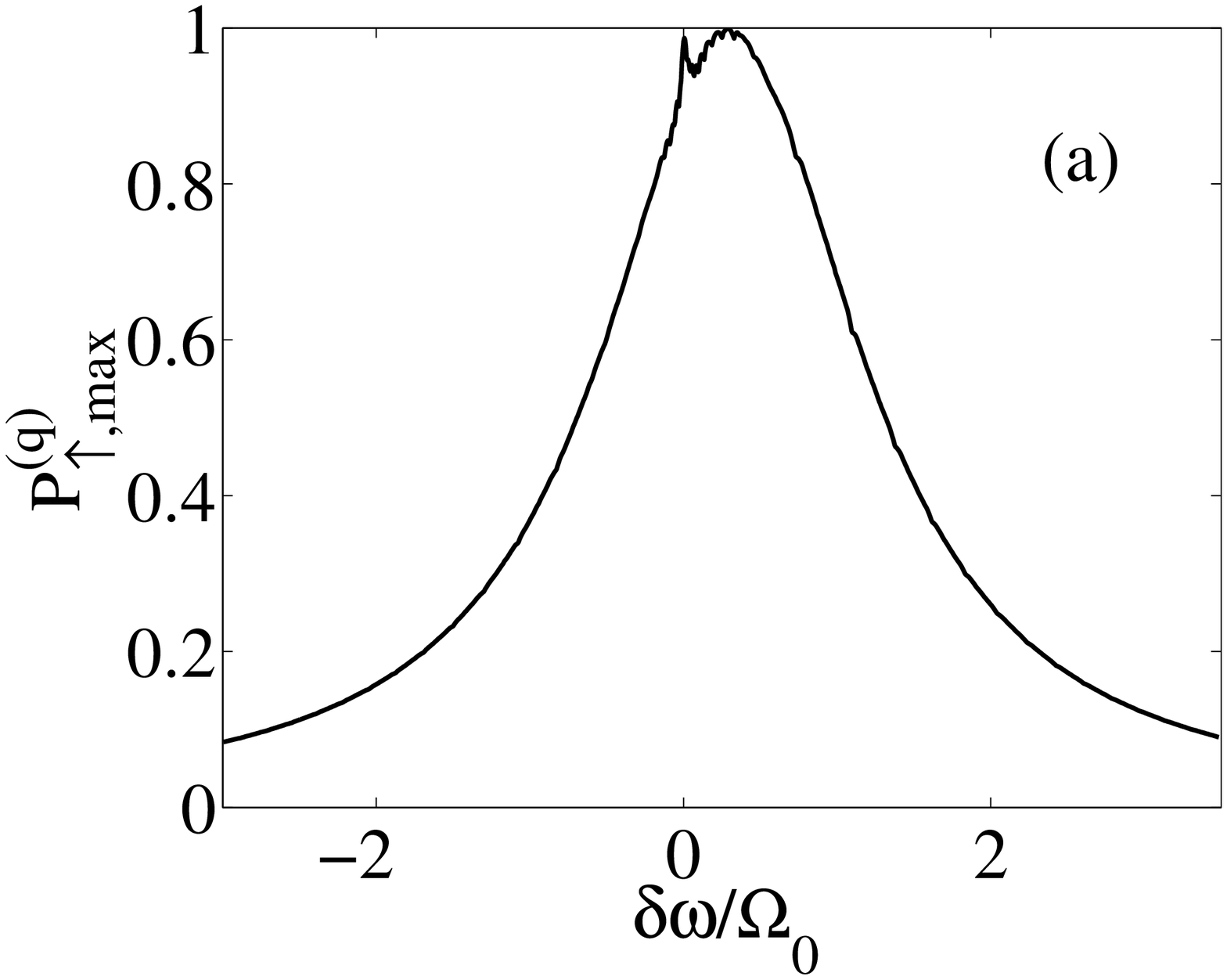}
\end{minipage}
\begin{minipage}[b]{5.3cm}
\includegraphics[width=5.3cm]{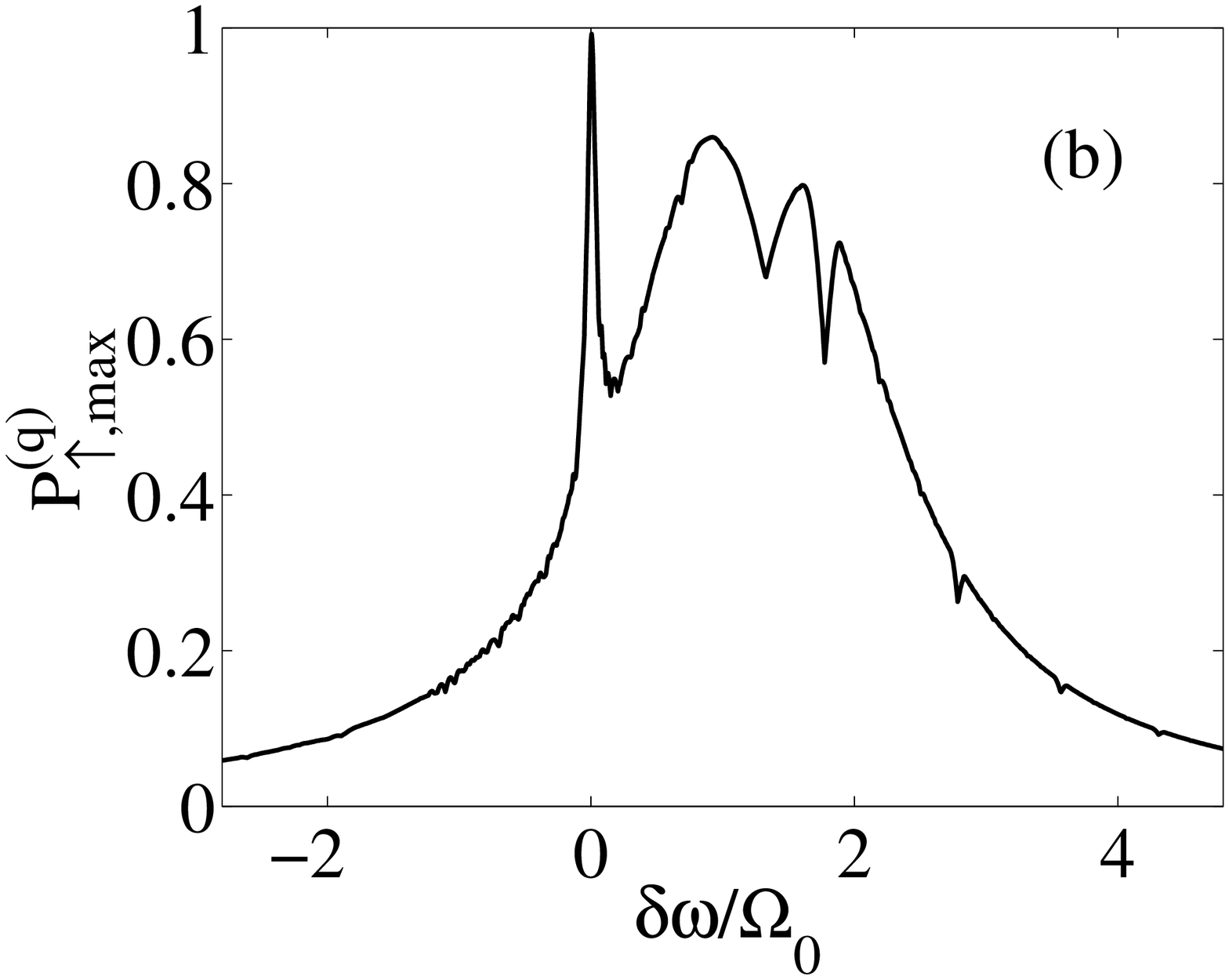}
\end{minipage}
\begin{minipage}[b]{5.3cm}
\includegraphics[width=5.3cm]{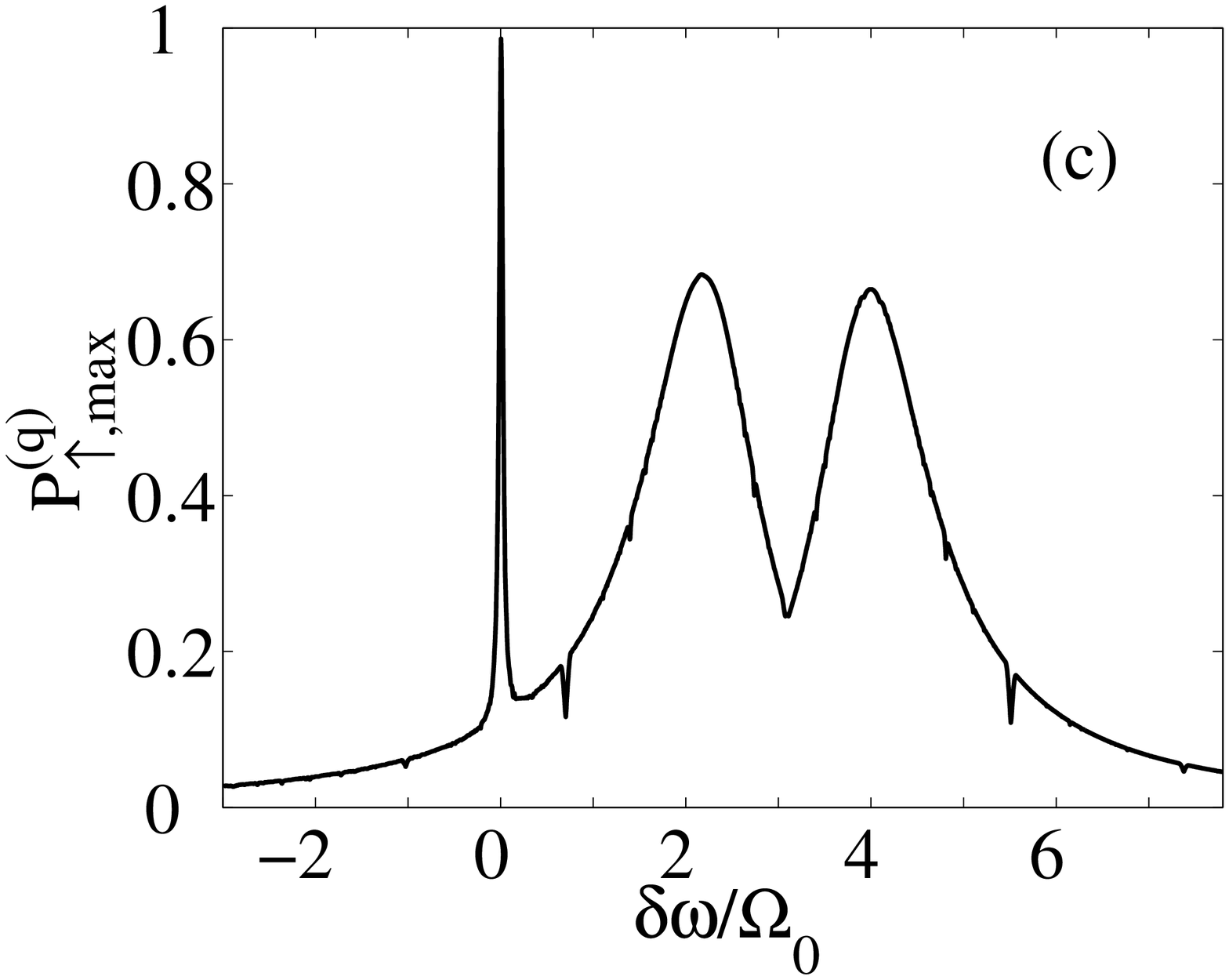}
\end{minipage}
\caption{Maximum qubit excitation probability P$_{\uparrow,{\rm
max}}^{(\rm q)}$ between $t=0$ and $t = 20\pi/\Omega_0$ for
$\lambda/\Omega_0 = $ 0.5 (a), 2 (b) and 5 (c). $\theta_{\rm
q}=\pi/4$, and $\theta_{\rm TLS}=\pi/6$.}
\end{figure}

\subsection{Experimental considerations}

In the early experiments on phase qubits coupled to TLSs
\cite{Simmonds,Cooper}, the qubit relaxation rate $\Gamma_1^{\rm
(q)}$ ($\sim$40MHz) was comparable to the splitting between the
two Rabi peaks $\lambda_{\rm ss}$ ($\sim$20-70MHz), whereas the
on-resonance Rabi frequency $\Omega_0$ was tunable from 30MHz to
400MHz. The constraint that $\Omega_0$ cannot be reduced to values
much lower than the decoherence rate made the strong-coupling
regime, where $\Omega_0\ll\lambda_{\rm ss}$, inaccessible.
Although the intermediate-coupling regime was accessible,
observation of the additional features in Fig. 2 discussed above
would have required a time at least comparable to the qubit
relaxation time. With the new qubit design of Ref.
\cite{Martinis2}, the qubit relaxation time has been increased by
a factor of 20. Therefore, all the effects that were discussed
above should be observable.

We finally consider one possible application of our results to
experiments on phase qubits, namely the problem of characterizing
an environment comprised of TLSs. Since measurement of the
locations of the three peaks in Fig. 2 provides complete
information about the four-level spectrum, both $\lambda_{\rm cc}$
and $\lambda_{\rm ss}$ can be extracted from such results. One can
therefore obtain the distribution of values of both $E_{\rm TLS}$
and $\theta_{\rm TLS}$ for all the TLSs in the environment. Note
that in some cases, e.g. a phase qubit coupled to the TLSs through
the operator of charge across the junction, we find that
$\theta_{\rm q}=\pi/2$, and therefore $\lambda_{\rm cc}$ vanishes
for all the TLSs. In that case the two-photon peak would always
appear at the midpoint (to a good approximation) between the two
main Rabi peaks. Although that would prevent the determination of
the values of $E_{\rm TLS}$ and $\theta_{\rm TLS}$ separately, it
would provide information about the qubit-TLS coupling mechanism.

\section{Conclusion}

We have studied the problem of a qubit that is coupled to an
uncontrollable two-level system and a background environment. We
have derived analytic expressions describing the contribution of a
quantum TLS to the qubit decoherence dynamics, and we have used
numerical calculations to test the validity of those expressions.
Our results can be considered a generalization of the well-known
results of the traditional weak-coupling approximation.
Furthermore, our results concerning the qubit's response to an
oscillating external field can be useful to experimental attempts
to characterize the TLSs surrounding a qubit, which can then be
used as part of possible techniques to eliminate the TLS's
detrimental effects on the qubit operation.

\section*{Acknowledgments}
This work was supported in part by the National Security Agency
(NSA) and Advanced Research and Development Activity (ARDA) under
Air Force Office of Research (AFOSR) contract number
F49620-02-1-0334; and also supported by the National Science
Foundation grant No.~EIA-0130383. One of us (S. A.) was supported
by a fellowship from the Japan Society for the Promotion of
Science (JSPS).


\begin{thebibliography}{99}
\bibitem{Assumptions} We shall make a number of assumptions that
might seem too specific at certain points. We have, however, been
careful not to make a choice of parameters that causes any of the
interesting physical phenomena to disappear. For a more detailed
discussion of our assumptions, see Ref. \cite{Ashhab2,Ashhab1}.
\bibitem{Temperature} Note that in this paper we shall only consider the
zero temperature case. For a treatment of the finite temperature
case, see Ref. \cite{Ashhab1}.
\end{thebibliography}
\end{document}